\documentclass[]{AO4ELT}  

\usepackage{microtype}
\usepackage{biblatex}
\usepackage{amsmath,amsfonts,amssymb}
\usepackage{graphicx}
\usepackage{pst-all} 
\usepackage[colorlinks=true, allcolors=blue]{hyperref}
\usepackage{graphicx}	
\usepackage{amsmath}	
\usepackage{comment}
\usepackage{xspace} 
\usepackage{mathtools}
\usepackage{hyperref}
\usepackage{float}
\usepackage{bm}
\usepackage{pdflscape}
\usepackage{pdfpages}
\usepackage{stmaryrd}
\usepackage{setspace}
\DeclareFontFamily{U}{mathx}{}
\DeclareFontShape{U}{mathx}{m}{n}{<-> mathx10}{}
\DeclareSymbolFont{mathx}{U}{mathx}{m}{n}
\DeclareMathAccent{\widecheck}{0}{mathx}{"71}

\usepackage{multirow} \usepackage{threeparttable}
\usepackage{booktabs} \usepackage{stfloats}

\newcommand*{\V}[1]{\boldsymbol{#1}}   
\newcommand*{\M}[1]{\mathbf{#1}}       
\newcommand*{\TransposeLetter}{\hspace*{-.25ex}\top\hspace*{-.25ex}}
\newcommand*{\T}{^{\TransposeLetter}} 
\DeclareFontFamily{U}{mathx}{\hyphenchar\font45}
\DeclareFontShape{U}{mathx}{m}{n}{<-> mathx10}{}
\DeclareSymbolFont{mathx}{U}{mathx}{m}{n}
\DeclareMathAccent{\widebar}{0}{mathx}{"73}

\DeclarePairedDelimiterX{\Paren}[1]{(}{)}{#1}
\DeclarePairedDelimiterX{\Brace}[1]{\{}{\}}{#1}
\DeclarePairedDelimiterX{\Brack}[1]{[}{]}{#1}
\DeclarePairedDelimiterX{\Abs}[1]{\rvert}{\lvert}{#1}
\DeclarePairedDelimiterX{\Norm}[1]{\lVert}{\rVert}{#1}
\DeclarePairedDelimiterX{\Avg}[1]{\langle}{\rangle}{#1}
\DeclarePairedDelimiterX{\Round}[1]{\lfloor}{\rceil}{#1}
\DeclarePairedDelimiterX{\Floor}[1]{\lfloor}{\rfloor}{#1}
\DeclarePairedDelimiterX{\Ceil}[1]{\lceil}{\rceil}{#1}
\DeclarePairedDelimiterX{\Inner}[2]{\langle}{\rangle}{#1,#2}

\DeclareMathOperator{\Trace}{tr}

\DeclarePairedDelimiterXPP{\Expect}[1]{\mathbb{E}}(){}{#1}

\newcommand*{\estim}[1]{\widehat{#1}}
\newcommand{\Sest}{\widehat{\M{S}}}

\usepackage{numprint}

\makeatletter
\def\widebreve{\mathpalette\wide@breve}
\def\wide@breve#1#2{\sbox\z@{$#1#2$}%
     \mathop{\vbox{\m@th\ialign{##\crcr
\kern0.08em\brevefill#1{0.8\wd\z@}\crcr\noalign{\nointerlineskip}%
                    $\hss#1#2\hss$\crcr}}}\limits}
\def\brevefill#1#2{$\m@th\sbox\tw@{$#1($}%
  \hss\resizebox{#2}{\wd\tw@}{\rotatebox[origin=c]{90}{\upshape(}}\hss$}
\makeatletter

\makeatletter         
\def\@maketitle{
\includegraphics[width = 170mm]{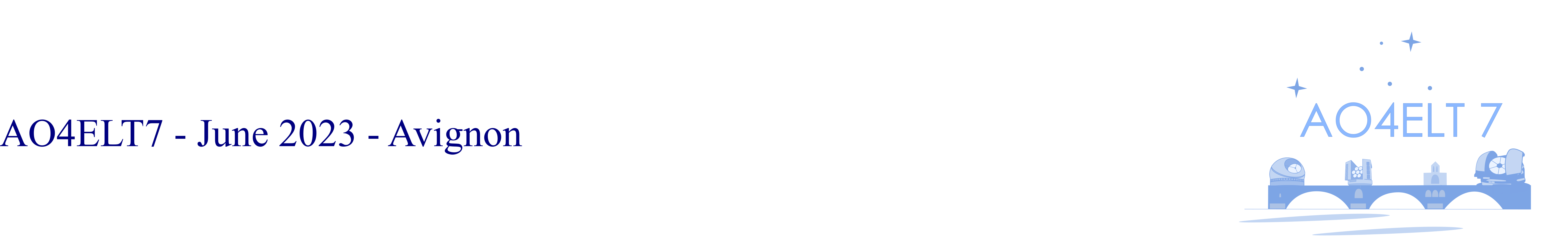}\\[8ex]
\begin{center}
{\Huge \bfseries \sffamily \@title }\\[4ex] 
{\Large  \@author}\\[4ex] 
\@date
\end{center}}

\title{Combining statistical learning with deep learning for improved exoplanet detection and characterization}

\author[a]{Olivier Flasseur}
\author[b]{Théo Bodrito}
\author[c]{Julien Mairal}
\author[b,d]{Jean Ponce}
\author[a]{Maud Langlois}
\author[e,f]{Anne-Marie Lagrange}
\affil[a]{Centre de Recherche Astrophysique de Lyon, CNRS, Univ. de Lyon, \newline Univ. Claude Bernard Lyon 1, ENS de Lyon, France}
\affil[b]{Département d'Informatique de l'{\'E}cole Normale Supérieure \newline (ENS-PSL, CNRS, Inria), France}
\affil[c]{Univ. Grenoble Alpes, Inria, CNRS, Grenoble INP, LJK, \newline 38000 Grenoble, France}
\affil[d]{Courant Institute of Mathematical Sciences, \newline Center for Data Science, New York Univ., USA}
\affil[e]{Laboratoire d'{\'E}tudes Spatiales et d'Instrumentation en Astrophysique, Observatoire de Paris, Univ. PSL, Sorbonne Univ., Univ. Paris Diderot, France}
\affil[f]{Univ. Grenoble Alpes, Institut de Planétologie \newline et d'Astrophysique de Grenoble, France}

\authorinfo{Further author information: Send correspondence to Olivier Flasseur.\\E-mail: olivier.flasseur@univ-lyon1.fr, Telephone: +33 (0)4 78 86 83 70.}

\begin{document} 
\maketitle
\begin{abstract}
In direct imaging at high contrast, the bright glare produced by the host star makes the detection and the characterization of sub-stellar companions particularly challenging. In spite of the use of an extreme adaptive optics system combined with a coronagraphic mask to strongly attenuate the starlight contamination, dedicated post-processing methods combining several images recorded with the pupil tracking mode of the telescope are needed to reach the required contrast.

Among the large variety of post-processing methods of the field, the PACO algorithm capturing locally the spatial correlations of the data with a multi-variate Gaussian model shown better detection sensitivity than standard methods of the field (e.g., cADI, PCA, TLOCI). However, there is a room for improvement to increase the detection sensitivity due to the approximate fidelity of the  statistical model embedded in PACO with respect to the observations.

In that context, we recently proposed to combine the statistics-based model of PACO with a deep learning approach in a three-step algorithm. First, the data are centered and whitened locally using the PACO framework to improve the stationarity and the contrast in a preprocessing step. Second, a convolutional neural network (CNN) is trained in a supervised fashion to detect the signature of synthetic sources in the preprocessed science data. Finally, the trained network is applied to the preprocessed observations and delivers a detection map. A second network is trained to infer locally the photometry of detected sources. Both deep models are trained from scratch with a custom data augmentation strategy allowing to generate a large training set from a single spatio-temporo-spectral dataset. This strategy can be applied to process jointly the images of observations conducted with angular, and eventually spectral, differential imaging (A(S)DI). In this proceeding, we present in a unified framework the key ingredients of the deep PACO algorithm both for ADI and ASDI.

We apply our method on several datasets from the the IRDIS imager of the VLT/SPHERE instrument. Our method reaches, in average, a better trade-off between precision and recall than the comparative algorithms.
\end{abstract}


\section{INTRODUCTION}
\label{sec:introduction}

The future thirty meters class telescopes (e.g., ELT, GMT, TMT) will enable exploring much deeper the inner environment of nearby solar-type stars than existing facilities. This goal raises several challenges from a data science point of view, including: (i) approaching the ultimate performance of the instruments by an optimal extraction of the signals of the sought objects, (ii) capturing a highly spatially structured nuisance component subject to strong temporal fluctuations, and (iii) building a model of the nuisance component from several datasets to bypass the limits of ADI at very short angular separations. In that context, data-driven approaches combining statistical modeling with deep learning could be highly valuable to deal with the complexity of high-contrast observations.

In previous works, we have developed the PACO algorithm \cite{flasseur2018exoplanet,flasseur2018unsupervised,flasseur2018SPIE,flasseur2020robust,flasseur2020pacoasdi} dedicated to the post-processing of A(S)DI observations \cite{marois2006angular,racine1999speckle} for exoplanet detection and characterization by direct imaging at high contrast. PACO captures locally the spatial correlations of the nuisance component (i.e., speckles plus other sources of noise) with a scaled mixture of multi-variate Gaussian models. Its parameters are estimated in a data-driven fashion at the scale of a patch of a few tens of pixels. PACO delivers reliable detection confidences with an improved sensitivity with respect to the classical processing methods of the field (e.g., cADI \cite{marois2006angular,lagrange2009probable}, PCA \cite{soummer2012detection,amara2012pynpoint}, TLOCI \cite{marois2013tloci}). However, it remains room for improvement, especially at short angular separations. 

Very recently, we proposed a new algorithm, dubbed deep PACO \cite{flasseur2022exoplanet,flasseur2023deeppaco}, improving the detection sensitivity of PACO. The proposed method combines the statistics-based model of PACO with a deep learning model in a three steps procedure. First, the data are centered and whitened locally using the PACO framework to improve the stationarity and the contrast in a pre-processing step. Second, a convolutional neural network is trained from scratch, in a supervised fashion, to detect the signature of synthetic sources in the pre-processed science data. Finally, the trained network is applied to the pre-processed observations and delivers a detection map. Photometry of detected sources can be estimated by a second deep neural network. Both models are trained from scratch with a custom data augmentation strategy allowing to generate large training sets from a single spatio-temporo-spectral dataset. 

We review the main ingredients of the deep PACO algorithm in Sect. \ref{sec:algorithm}. Section \ref{sec:results} presents comparative results obtained on VLT/SPHERE data. Finally, Sect. \ref{sec:conclusion} draws our main conclusions and briefly discusses some future research directions. 
 
\section{THE DEEP PACO ALGORITHM}
\label{sec:algorithm}

\subsection{Direct Model of the Observations}
\label{subsec:direct_model_observations}

A stack $\V r \in \mathbb{R}^{N \times T \times L}$ of high-contrast images is composed of $N$-pixels frames, recorded at times $t \in \llbracket 1 ; T \rrbracket$ and in spectral channels $\ell \in \llbracket 1 ; L \rrbracket$. The contribution $\V r_{\ell} \in \mathbb{R}^{N \times T}$ writes: 
\begin{equation}
	\V r_\ell = \V f_\ell + \sum\limits_{p=1}^P \alpha_{p,\ell} \, \V h_\ell (\phi_p)\,,
	\label{eq:image_formation_model}
\end{equation}
where $\V f_\ell \in \mathbb{R}^{N \times T}$ and $\V h_\ell \in \mathbb{R}^{N \times T}$ are respectively the contribution of the nuisance component (i.e., speckles and other additive sources of noise) and of any point-like source taking the form of the off-axis point-spread function (PSF), at spectral channel $\ell$. The contribution of a source $p \in \llbracket 1 ; P \rrbracket$ is centered at location $\mathcal{F}_{t,\ell}(\phi_p)$ in the $t$-th image of the $\ell$-th channel, where $\phi_p$ is its initial location on an image at a time $t_{\text{ref}}$ and spectral channel $\lambda_{\text{ref}}$ of reference. $\mathcal{F}_{t,\ell}$ is a deterministic function accounting for the apparent motion of the sought sources induced by ASDI.
  
\subsection{Statistical Modeling of the Nuisance Component}
\label{subsec:statistical_learning}

As in \cite{flasseur2020pacoasdi}, we model the fluctuations of the nuisance component $\V f$ by a statistical model whose parameters are estimated locally (in this work, at a scale of non-overlapping square patches of $K$ pixels). The locality of the model allows to account for the non-stationarity of the nuisance. We note $\mathbb{P}$ the set of pixel locations on which the parameters of the statistical model should be evaluated. The distribution of a patch of nuisance $\V f_{n, t, \ell}$ centered at pixel $n \in \mathbb{P}$, at time $t$, and channel $\ell$ is modeled by a scaled multi-variate Gaussian: $\mathcal{N}(\V m_{n,\ell}, \sigma_{n,t,\ell}^2\M C_n)$. The sample estimators $\lbrace \widehat{\V m}_{n,\ell}\,, \widehat{\sigma}_{n,t,\ell}^{2} \,, \widehat{\M S}_n\rbrace$ of $\lbrace \V m_{n,\ell}\,, \sigma_{n,t,\ell}^{2} \,, \M C_n\rbrace$ are obtained from the $T\,L$ patches $\V r_n \in \mathbb{R}^{K \times T \times L}$ through the maximum-likelihood:
\begin{equation}
	\begin{cases}
		\widehat{\V m}_{n,\ell} =  \big(\sum\limits_t \widehat{\sigma}_{n,t,\ell}^{-2}\big)^{-1} \sum\limits_{t} \widehat{\sigma}_{n,t,\ell}^{-2} \, \V r_{n,t,\ell}\,,\\
		\widehat{\M S}_n = \frac{1}{T\,L} \sum\limits_{t, \ell}  \widehat{\sigma}_{n,t,\ell}^{-2} (\V r_{n,t,\ell} - \widehat{\V m}_{n,\ell}) (\V r_{n,t,\ell} - \widehat{\V m}_{n,\ell})\T\,,\\
		\widehat{\sigma}_{n,t,\ell}^{2} = K^{-1} (\V r_{n,t,\ell} - \widehat{\V m}_{n,\ell})\T \, \widehat{\M S}_n^{-1} (\V r_{n,t,\ell} - \widehat{\V m}_{n,\ell})\,,
	\end{cases}
	\label{eq:sample_estimators}
\end{equation}
\noindent The number of samples involved in the computation of $\widehat{\M S}_n$ being lower than the number $K$ of pixels in a patch, the sample covariance $\widehat{\M S}_n$ is very noisy. As in \cite{flasseur2020pacoasdi}, we regularize it by \textit{shrinkage}  \cite{ledoit2004well,chen2010shrinkage} to form $\estim{\M C}_n$ as follows: 
\begin{equation}
	\estim{\M C}_n = (1 - \widehat{\rho}_n)\,\widehat{\M S}_n + \widehat{\rho}_n\,\widehat{\M F}_n\,,
	\label{eq:shrunk_covariances}
\end{equation}
where $\widehat{\M F}_n$ is a diagonal matrix with only the sample variances. The weight $\widehat{\rho}_n$ sets a bias-variance trade-off and it is estimated in data-driven fashion:
\begin{equation}
  \estim\rho_n= \frac{
    \Trace\Paren[\big]{\Sest_{n}^2}
    + \Trace^2\Paren[\big]{\Sest_{n}}
    - 2\sum_{k=1}^K\Brack[\big]{\Sest_{n}}_{kk}^2
  }{
    (Q+1)\,\Paren[\Big]{
      \Trace\Paren[\big]{\Sest_{n}^2}
      - \sum_{k=1}^K\Brack[\big]{\Sest_{n}}_{kk}^2
    }
  } \,,
   \label{eq:shrinkfactor}
\end{equation}
with $Q = \big( \sum_{t,\ell} \widehat{\sigma}_{n,t,\ell}^{-2} \big)^2 / \big( \sum_{t,\ell} \widehat{\sigma}_{n,t,\ell}^{-4} \big)$ the equivalent number of patches involved in the computation of $\widehat{\M C}_n$ in the presence of the scaling factors $\lbrace \widehat{\sigma}_{n,t,\ell}^2 \rbrace_{t=1:T, \ell=1:L}$. Given the statistics of the nuisance, the dataset $\V r$ is preprocessed by centering and whitening to form $\widetilde{\V r} \in \mathbb{R}^{N \times T \times L}$ with attenuated spatial structures:
\begin{equation}
	\widetilde{\V r}_{n,t,\ell} = \M W_{n,t,\ell} \, \V r_{n,t,\ell} = \widehat{\sigma}_{n,t,\ell}^{-1} \, \widehat{\M L}_{n}\T \, \left( \V r_{n,t,\ell} - \widehat{\V m}_{n,\ell} \right) \,,
	\label{eq:whitening}
\end{equation}
such that $\widehat{\M L}_n \widehat{\M L}_n\T = \widehat{\M C}_n^{-1}$.

\subsection{Detection by Supervised Deep Learning}
\label{subsec:detection_supervised_deep_learning}


We aim to produce a detection map $\widehat{\V y}$ in $\left[0 ; 1 \right]^{M}$, where each pixel-value represents the pseudo-probability that a source is centered at that location at time $t_\text{ref}$. In order to perform supervised training of our model, we generate $S$ pairs $\lbrace \widebreve{\V r}^{[s]} ; \V y^{[s]} \rbrace_{s=1:S}$ of synthetic samples/ground truths, resulting from the massive injection of point-like sources. The trained model is data-dependent, i.e. it differs for each A(S)DI dataset due to the high variability of the nuisance component from one observation to the other. We thus resort to data-augmentation to generate a large training basis from a single dataset: we apply a random permutation of the $T$ images of each spectral channel for each training sample $s$. 
Synthetic sources are injected inside the temporally permuted data following the direct model (\ref{eq:image_formation_model}) to form the intermediate datasets $\lbrace \widebar{\V r} \in \mathbb{R}^{N \times T \times L} \rbrace_{s=1:S}$:
\begin{equation}
	\widebar{\V r}_\ell^{[s]} = \M P^{[s]} \, \V f_\ell + \sum\limits_{p=1}^{P^{[s]}} \alpha_{p,\ell}^{[s]} \, \V h_\ell (\phi_p^{[s]})\,,
	\label{eq:injection_model}
\end{equation}
with $\M P$ the operator performing a random permutation of the temporal frames. 
For each synthetic source $p$ and if $L > 2$, we select randomly a synthetic SED $\V \alpha_p \in \mathbb{R}^L$ from a custom library of 10,000 sub-stellar spectra generated with ExoREM \cite{charnay2018self}.
Before injection of synthetic sources, the initial dataset $\V r$ is pre-processed to form $\widetilde{\V r}$. After injection of synthetic sources, the set $\mathbb{S}^{[s]}$ of locations impacted by the signal of the fiducial sources is determined.
Outside $\mathbb{S}^{[s]}$, 
the pre-processed images are obtained from the temporal permutation of $\widetilde{\V r}$.
Inside $\mathbb{S}^{[s]}$, the statistics of the nuisance and the pre-processed images are updated given the contamination of the $P^{[s]}$ injected sources to form $\lbrace \widecheck{\V r} \in \mathbb{R}^{N \times T \times L} \rbrace_{s=1:S}$ such that:
\begin{equation}
	\widecheck{\V r}_{n,t,\ell}^{[s]} = \begin{cases}
  \M W_{n,t,\ell} \, \widebar{\V r}_{n,t,\ell}^{[s]}, & \text{for } n \in \mathbb{S}^{[s]}\, \cap\, \mathbb{P}\,, \\
	\M P^{[s]} \, \widetilde{\V r}_{n,t,\ell}, & \text{for } n \in \mathbb{P} - \mathbb{S}^{[s]}\, \cap\, \mathbb{P}\,.
\end{cases}
	\label{eq:preprocessing_local_update}
\end{equation}
Finally, the apparent motion of the injected sources are compensated to co-align their signals within $\lbrace \widebreve{\V r}^{[s]} \in \mathbb{R}^{N \times T \times L} \rbrace_{s=1:S}$:
\begin{equation}
	\widebreve{\V r}_{t,\ell}^{[s]} = \M D_{t,\ell} \, \widecheck{\V r}_{t,\ell}^{[s]}\,,
	\label{eq:preprocessing_derotation}
\end{equation} 
with $\M D_{t,\ell}$ the operator performing a rotation of the image at time $t$ and spectral channel $\ell$ by the opposite of the parallactic angle at time $t$.


At training time, the network parameters are optimized by minimizing the Dice2 loss \cite{wang2020improved}:
\begin{equation}
	\mathcal{L}^{[s]}  = 1 - \frac{\sum\limits_{m} \V y_m^{[s]} \, \widehat{\V y}_m^{[s]} + \epsilon}{\sum\limits_{m} \V y_m^{[s]} + \widehat{\V y}_m^{[s]} + \epsilon}\, - \,\frac{\sum\limits_{m} (1-\V y_m^{[s]})(1-\widehat{\V y}_m^{[s]} + \epsilon)}{\sum\limits_{m} 2 - \V y_m^{[s]} - \widehat{\V y}_m^{[s]} + \epsilon}\,,
	\label{eq:loss}
\end{equation}
where $\lbrace  \V y^{[s]} ; \widehat{\V y}^{[s]} \rbrace$ is a set of ground truth and predicted detection maps, and $\epsilon$ is a small stability parameter. At validation time, we compute from a predicted detection map $\widehat{\V y}^{[s]} \in \left[ 0 ; 1 \right]^M$ thresholded at $\tau \in \left[0 ; 1\right]$, the true positive rate (TPR) and the false discovery rate (FDR). 
From these two quantities, receiver operating curves (ROCs) are built and the area under ROCs (AUC) is derived as an overall measurement of the trade-off between precision and recall. 

Concerning the architecture, we use a U-Net with a ResNet18 as encoder backbone\footnote{\href{https://github.com/qubvel/segmentation\_models.pytorch}{https://github.com/qubvel/segmentation\_models.pytorch}}. The network weights are optimized from scratch with AMSGrad. The number of epochs, the number of samples per epoch, the batch size, the weight decay, and the learning rate are discussed in details in \cite{flasseur2022exoplanet, flasseur2023deeppaco}.

\section{RESULTS ON VLT/SPHERE DATA}
\label{sec:results}

\begin{figure}[t]
	\begin{center}
		\includegraphics[width=\textwidth]{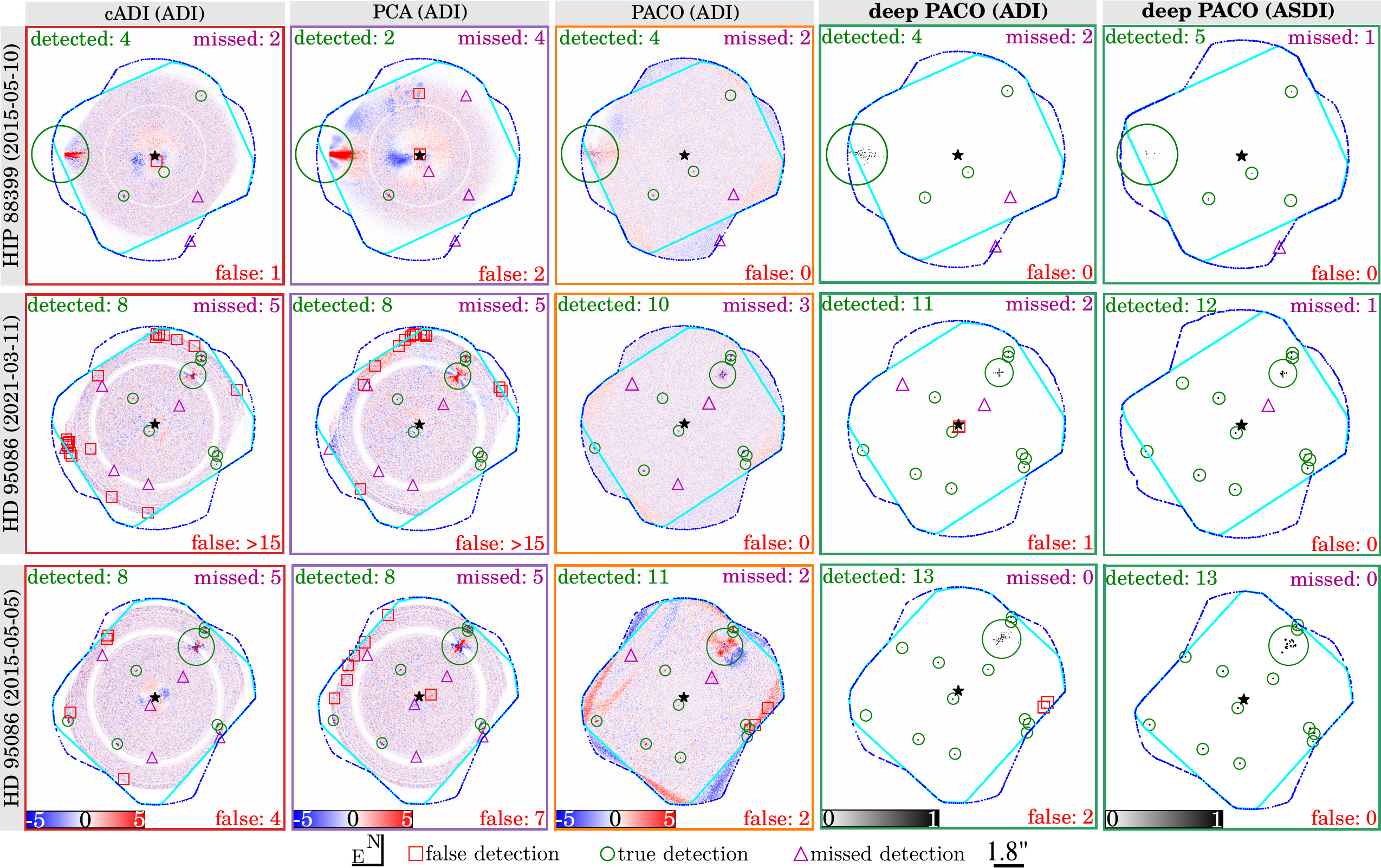}
		\caption{Detection maps obtained with cADI, PCA, PACO, and deep PACO on three datasets (HIP 88399 --2015-05-10--, HD 95086 --2021-03-11--, HD 95086 --2015-05-05--) from the VLT/SPHERE-IRDIS instrument.}
		\label{fig:irdis_adi_vs_asdi_extended}
	\end{center}
\end{figure}

\noindent For our experiments, we have selected three datasets from the VLT/SPHERE-IRDIS imager \cite{beuzit2019sphere}. The performance of deep PACO are compared with the cADI, PCA, and PACO algorithms. The details about the observing parameters, the pre-reduction of the recorded datasets, and the setting of the comparative algorithms are discussed in \cite{flasseur2022exoplanet,flasseur2023deeppaco}.

Figure \ref{fig:irdis_adi_vs_asdi_extended} shows the detection maps obtained with the comparative algorithms. The detected peaks are classified as true, missed, or false detections based on a thresholding of the output map at $\tau=5$ for the algorithms producing a signal-to-noise map (i.e., cADI, PCA, PACO), and at $\tau = 0.5$ for deep PACO producing a pseudo-probability map. For the four considered algorithms, the results are obtained by processing the first spectral channel, i.e. exploiting solely ADI. For the proposed approach, we compare ADI and ASDI by also considering a joint processing of the two spectral channels of the dual band observations. These results illustrate the ability of deep PACO to reach, in average, a better detection sensitivity and, at the same time, to avoid some spurious false alarms produced by the comparative algorithms. The joint spectral processing allows to further improve the trade-off between precision and recall. 

\section{CONCLUSION}
\label{sec:conclusion}

As a proof of concept, we applied our method on several datasets from the VLT/SPHERE-IRDIS instrument. We compared the deep PACO  method against some state-of-the-art algorithms of the field, including PACO. With ADI, our method reaches in average a better trade-off between precision and recall than the comparative algorithms. A joint processing of spatio-temporo-spectral observations obtained with ASDI allows to further improve the detection performance.

On the methodological side, we are currently investigating ways to improve the deep learning stage of the proposed algorithm by jointly modeling the typical distribution of the nuisance component from multiple datasets. This methodology would strongly differs from reference differential imaging (RDI, \cite{ruane2019reference}) in the sense that a highly non-linear model would be learned from the archived observations and would exploit several prior domain knowledge. On the application side, we will consider  to adapt and to apply the deep PACO algorithm on simulated ELT high-contrast data (e.g., from ELT/HARMONI) to ground its benefits in the context of the future giant telescopes. 

\acknowledgments

This work is supported by the French National Programs (PNP and PNPS) and by the Action Spécifique Haute Résolution Angulaire (ASHRA) of CNRS/INSU co-funded by CNES. This project is also supported in part by the European Research Council (ERC) under the European Union's Horizon 2020 research and innovation programme (COBREX; grant agreement n° 885593). The work of TB and JP was supported in part by the Inria/NYU collaboration, the Louis Vuitton/ENS chair on artificial intelligence and the French government under management of Agence Nationale de la Recherche as part of the \textit{Investissements d'avenir} program, reference ANR19-P3IA0001 (PRAIRIE 3IA Institute). The work of JM was supported in part by the ERC grant number 714381 (SOLARIS project) and by ANR 3IA MIAI@Grenoble-Alpes (ANR-19-P3IA-0003).

\sloppy The raw data used in this article are freely available on the ESO archive facility at \href{http://archive.eso.org/eso/eso\_archive\_main.html}{http://archive.eso.org/eso/eso\_archive\_main.html}. They were pre-reduced with the SPHERE Data Centre, jointly operated by OSUG/IPAG (Grenoble), PYTHEAS/LAM/CESAM (Marseille), OCA/Lagrange (Nice), Observatoire de Paris/LESIA (Paris), and Observatoire de Lyon/CRAL (Lyon, France). This work was also granted access to the HPC resources of IDRIS under the allocation 2022-AD011013643 made by GENCI.

\printbibliography 

@inproceedings{flasseur2022exoplanet,
  title={Exoplanet detection in angular differential imaging: combining a statistics-based learning with a deep-based learning for improved detections},
  author={Flasseur, Olivier and Bodrito, Th{\'e}o and Mairal, Julien and Ponce, Jean and Langlois, Maud and Lagrange, Anne-Marie},
  booktitle={Adaptive Optics Systems VIII},
  volume={12185},
  pages={1154--1167},
  year={2022},
  organization={SPIE}
}

@article{flasseur2023deeppaco,
  title={deep PACO: Combining statistical models with deep learning for exoplanet detection and characterization in direct imaging at high contrast},
  author={Flasseur, Olivier and Bodrito, Th{\'e}o and Mairal, Julien and Ponce, Jean and Langlois, Maud and Lagrange, Anne-Marie},
  journal={under minor review in Monthly Notices of the Royal Astronomical Society},
  year={2023},
}

@article{charnay2018self,
  title={A self-consistent cloud model for brown dwarfs and young giant exoplanets: comparison with photometric and spectroscopic observations},
  author={Charnay, Benjamin and B{\'e}zard, Bruno and Baudino, J-L and Bonnefoy, Micka{\"e}l and Boccaletti, Anthony and Galicher, Rapha{\"e}l},
  journal={The Astrophysical Journal},
  volume={854},
  number={2},
  pages={172},
  year={2018},
  publisher={IOP Publishing}
}

@article{wang2020improved,
  title={An improved {D}ice loss for pneumothorax segmentation by mining the information of negative areas},
  author={Wang, Lu and Wang, Chaoli and Sun, Zhanquan and Chen, Sheng},
  journal={IEEE Access},
  volume={8},
  pages={167939--167949},
  year={2020},
  publisher={IEEE}
}

@article{ruane2019reference,
  title={Reference star differential imaging of close-in companions and circumstellar disks with the, {NIRC}2 vortex coronagraph at the {W}.{M}. {K}eck observatory},
  author={Ruane, Garreth and Ngo, Henry and Mawet, Dimitri and Absil, Olivier and Choquet, {\'E}lodie and Cook, Therese and Gonzalez, Carlos Gomez and Huby, Elsa and Matthews, Keith and Meshkat, Tiffany and others},
  journal={The Astronomical Journal},
  volume={157},
  number={3},
  pages={118},
  year={2019},
  publisher={IOP Publishing}
}

@article{amara2012pynpoint,
  title={{PYNPOINT}: an image processing package for finding exoplanets},
  author={Amara, Adam and Quanz, Sascha P},
  journal={Monthly Notices of the Royal Astronomical Society},
  volume={427},
  number={2},
  pages={948--955},
  year={2012},
  publisher={Blackwell Science Ltd Oxford, UK}
}

@article{beuzit2019sphere,
  title={{SPHERE}: the exoplanet imager for the {V}ery {L}arge {T}elescope},
  author={Beuzit, J-L and Vigan, Arthur and Mouillet, David and Dohlen, Kjetil and Gratton, Raffaele and Boccaletti, Anthony and Sauvage, J-F and Schmid, Hans Martin and Langlois, Maud and Petit, Cyril and others},
  journal={Astronomy \& Astrophysics},
  volume={631},
  pages={A155},
  year={2019},
  publisher={EDP Sciences}
}

@inproceedings{flasseur2018SPIE,
  title={Exoplanet detection in angular and spectral differential imaging: local learning of background correlations for improved detections},
  author={Flasseur, Olivier and Denis, Lo{\"\i}c and Thi{\'e}baut, {\'E}ric M and Langlois, Maud},
  booktitle={SPIE Astronomical Telescopes + Instrumentation},
  volume={10703},
  pages={107032R},
  year={2018},
  organization={International Society for Optics and Photonics}
}

@inproceedings{flasseur2018unsupervised,
  title={An unsupervised patch-based approach for exoplanet detection by direct imaging},
  author={Flasseur, Olivier and Denis, Loic and Thi{\'e}baut, {\'E}ric and Langlois, Maud},
  booktitle={IEEE International Conference on Image Processing},
  pages={2735--2739},
  year={2018},
}

@article{flasseur2018exoplanet,
  title={Exoplanet detection in angular differential imaging by statistical learning of the nonstationary patch covariances - {T}he {PACO} algorithm},
  author={Flasseur, Olivier and Denis, Lo{\"\i}c and Thi{\'e}baut, {\'E}ric and Langlois, Maud},
  journal={Astronomy \& Astrophysics},
  volume={618},
  pages={A138},
  year={2018},
  publisher={EDP Sciences}
}

@article{flasseur2020robust,
  title={Robustness to bad frames in angular differential imaging: a local weighting approach},
  author={Flasseur, Olivier and Denis, Lo{\"\i}c and Thi{\'e}baut, {\'E}ric and Langlois, Maud},
  journal={Astronomy \& Astrophysics},
  volume={634},
  pages={A2},
  year={2020},
  publisher={EDP Sciences}
}

@article{flasseur2020pacoasdi,
  title={PACO ASDI: an algorithm for exoplanet detection and characterization in direct imaging with integral field spectrographs},
  author={Flasseur, Olivier and Denis, Lo{\"\i}c and Thi{\'e}baut, {\'E}ric and Langlois, Maud},
  journal={Astronomy \& Astrophysics},
  volume={637},
  pages={A9},
  year={2020},
  publisher={EDP Sciences}
}

@article{marois2013tloci,
  title={TLOCI: A fully loaded speckle killing machine},
  author={Marois, Christian and Correia, Carlos and V{\'e}ran, Jean-Pierre and Currie, Thayne},
  journal={Proceedings of the International Astronomical Union},
  volume={8},
  number={S299},
  pages={48--49},
  year={2013},
  publisher={Cambridge University Press}
}

@article{soummer2012detection,
  title={Detection and characterization of exoplanets and disks using projections on {K}arhunen-{L}o{\`e}ve eigenimages},
  author={Soummer, R{\'e}mi and Pueyo, Laurent and Larkin, James},
  journal={The Astrophysical Journal Letters},
  volume={755},
  number={2},
  pages={L28},
  year={2012},
  publisher={IOP Publishing}
}

@article{marois2006angular,
  title={Angular Differential Imaging: A Powerful High-Contrast Imaging Technique},
  author={Marois, Christian and Lafreni{\`e}re, David and Doyon, Rene and Macintosh, Bruce and Nadeau, Daniel},
  journal={The Astrophysical Journal},
  volume={641},
  number={1},
  pages={556},
  year={2006},
  publisher={IOP Publishing}
}

@article{lagrange2009probable,
  title={A probable giant planet imaged in the $\beta$ {P}ictoris disk: {VLT}/{N}a{C}o deep {L}'-band imaging},
  author={Lagrange, A-M and Gratadour, D and Chauvin, G and Fusco, T and Ehrenreich, D and Mouillet, D and Rousset, G and Rouan, D and Allard, F and Gendron, {\'E} and others},
  journal={Astronomy \& Astrophysics},
  volume={493},
  number={2},
  pages={L21--L25},
  year={2009},
  publisher={EDP Sciences}
}

@article{racine1999speckle,
  title={Speckle noise and the detection of faint companions},
  author={Racine, Ren{\'e} and Walker, Gordon AH and Nadeau, Daniel and Doyon, Ren{\'e} and Marois, Christian},
  journal={Publications of the Astronomical Society of the Pacific},
  volume={111},
  number={759},
  pages={587},
  year={1999},
  publisher={IOP Publishing}
}

@article{ledoit2004well,
  title={A well-conditioned estimator for large-dimensional covariance matrices},
  author={Ledoit, Olivier and Wolf, Michael},
  journal={Journal of Multivariate Analysis},
  volume={88},
  number={2},
  pages={365--411},
  year={2004},
  publisher={Elsevier}
}

@article{chen2010shrinkage,
  title={Shrinkage algorithms for {MMSE} covariance estimation},
  author={Chen, Yilun and Wiesel, Ami and Eldar, Yonina C and Hero, Alfred O},
  journal={IEEE Transactions on Signal Processing},
  volume={58},
  number={10},
  pages={5016--5029},
  year={2010},
  publisher={IEEE}
}

\end{document}